\documentclass[aip,jcp,reprint]{revtex4-1}

\usepackage{graphicx}
\usepackage{amsmath,amsxtra,amssymb,latexsym, amscd,color}

\begin{document}

\title{Folding and escape of nascent proteins at ribosomal exit tunnel}

\author{Bui Phuong Thuy}
\affiliation{Center for Computational Physics,
Institute of Physics, Vietnam Academy of Science and Technology, 10 Dao Tan, Ba
Dinh, Hanoi, Vietnam}
\affiliation{Nam Dinh University of Technology Education, Phu Nghia, Loc Ha,
Nam Dinh, Vietnam}

\author{Trinh Xuan Hoang}
\email{hoang@iop.vast.ac.vn}
\affiliation{Center for Computational Physics,
Institute of Physics, Vietnam Academy of Science and Technology,
10 Dao Tan, Ba Dinh, Hanoi, Vietnam}

\date{February 2016}

\begin{abstract}
We investigate the interplay between post-translational
folding and escape of two small single-domain proteins at the ribosomal exit
tunnel by using Langevin
dynamics with coarse-grained models. It is shown that at temperatures lower or
near the temperature of the fastest folding, folding proceeds concomitantly
with the escape process, resulting in vectorial folding and enhancement of
foldability of nascent proteins. The concomitance between the two
processes, however, deteriorates as temperature increases. Our folding
simulations as well as free energy
calculation by using umbrella sampling show that, at low temperatures, folding
at the tunnel follows one or two specific pathways without kinetic traps. 
It is shown that the
escape time can be mapped to a one-dimensional diffusion model with
two different regimes for temperatures above and below the folding
transition temperature. Attractive interactions between amino
acids and attractive sites on the tunnel wall lead to a free energy barrier
along the escape route of protein. It is suggested that this barrier slows
down the escape process and consequently promotes correct folding of
the released nascent protein.
\end{abstract}

\maketitle

\section{Introduction}

A majority of nascent proteins must be folded into their native states shortly
after emerging from the ribosome in order to avoid aggregation in the 
complex and crowded environment of the cell \cite{Dobson2003}. This task
is partially fulfilled thanks to a variety of molecular chaperones
that assist folding, as well as help preventing and
repairing misfolded proteins \cite{Frydman}. Another factor, that has
been under intense research, is the effects of the ribosome on folding of
nascent proteins \cite{Dobson2010,Baldwin,Deutsch2005,Deutsch2009,Bustamante,Ugrinov,Evans,Sanchez}. 
Several studies have demonstrated that the ribosome facilitates correct
folding of nascent proteins \cite{Bustamante,Ugrinov}.
The effects of the ribosome on nascent protein folding could be classified as
due to two reasons: a) the elongation of polypeptide chain during translation
and b) the confinement of ribosomal exit tunnel and its interactions with
nascent proteins. The first one is associated with the hypothesis of
cotranslational protein folding, namely folding that occurs during protein
synthesis, which receives now a substantial experimental support
\cite{Deutsch2005,Deutsch2009}.
A basic feature of cotranslational protein folding is vectorial folding
\cite{Baldwin}, which begins from the N-terminus and proceeds to the C-terminus
of protein, as a result of vectorial synthesis. 
This folding mechanism is quite different from that of refolding
\cite{Anfinsen} of a free denatured protein in solution.
The impact of vectorial folding on folding efficiency is expected to be much
stronger in multi-domain proteins than in single domain proteins as it
minimizes the chance of interdomain misfolding \cite{Frydman99}.  Another
feature of cotranslational protein folding is that it is influenced by
translation rates, both locally \cite{Zhang} and globally \cite{Siller}. It has
been shown that changing local translation rates by using alternative
RNA codons leads to a significant effect \cite{EPO1}, and may
coordinate cotranslational folding \cite{EPO2}. 
On the other hand, protein structures may have evolved to exploit the effects
of cotranslational folding.  It has been found that the propensity to form an
$\alpha$-helix, obtained by analysis of PDB structures, is significantly higher
near the C-terminus than near the N-terminus of proteins, in agreement with a
non-equilibrium effect found in the folding of a growing chain \cite{HoangPRL}. 

The ribosomal polypeptide exit tunnel \cite{Voss} is a narrow passage, through
which nascent proteins are secreted into cellular environment. Originated at
the peptidyl transferase center (PTC), where peptide bond formation
takes place, the tunnel traverses through the body of ribosome large unit, and
exits at the ribosome outer surface on the opposite side. It is constructed by
aligned segments of the 23S rRNA molecule and several ribosomal proteins 
\cite{Pande2008}. 
The length of ribosomal tunnel is $\sim$100 $\AA$, whereas its diameter varies
in a range of approximately 10--20 $\AA$ \cite{Deutsch2009}.  The tunnel is
narrowest and also slightly bended at the constriction site of about 20 $\AA$
distal from the PTC.  The narrow confinement of the tunnel would hinder the
formation of tertiary structures \cite{Voss}, whereas simple structural units
such as $\alpha$-helices and $\beta$-hairpin have been reported to form inside
the tunnel \cite{Deutsch2005,Deutsch2009}.  Experimental studies also have
indicated that the ribosomal tunnel is a peptide-sensitive tunnel
\cite{Ito2002} and may play an active role in modulation of translation and
folding of nascent
proteins \cite{Ito2004,Dresios}. Simulations of cotranslational protein folding
revealed closer details of the folding process at the exit tunnel, and
suggest that the impact of the tunnel on folding is rather not universal, but
secondary structure dependent \cite{Thirumalai,Makarov} and protein dependent
\cite{Elcock,Marek}. 

In this study, we investigate the effects of ribosomal exit tunnel on folding
of nascent proteins with a focus on what going on after the translation is
complete and the polypeptide chain is no longer bound to the PTC. In contrast
to the growth of the polypeptide chain during translation, which is, to some
extent, quite steady and predetermined, the subsequent process in which the
full length protein escapes from the exit tunnel is stochastic with much less
controls by the ribosome. We are interested in how the confined geometry of the
exit tunnel and its interaction with nascent proteins influence their escape
and folding processes.  The aim is also to elucidate the interplay between
these two processes and its dependence on temperature.  With an attempt to map
the dynamics of the escape process to one-dimensional diffusion model, we would
like to make the problem general and predictable.  We will confine ourself to
an idealized model of the ribosomal tunnel, in the form of a hollow cylinder,
with an assessment that the detailed shape of the tunnel also plays some role
but is less substantial.  The proteins are considered in a coarse-grained
Go-like model \cite{Go}. These simplified models and molecular dynamics method
combined with umbrella sampling \cite{Torrie} allow us to obtain a
comprehensive picture of the non-equilibrium characteristics of the folding and
the escape processes, as well as equilibrium characteristics such as effective
free energy along the tunnel axis. We will consider two small single-domain
proteins of approximately the same size but different native state topology for
comparison.  The interaction of the tunnel with proteins will be modeled by
putting attractive sites on the tunnel wall with a potential homologous for all
amino acids, as a first order approximation.

The remaining of the paper is organized as follows. Section 2 provides a
detailed description of the coarse-grained models used for the proteins and for
the exit tunnel. Section 3 provides the results on folding and escape of
nascent proteins at the exit tunnel along with a diffusion description for the
escape process and the effects of attractive sites. Section 4 contains
a discussion of the results. Section 5 provides a conclusion. Details of the
simulation methods are given in Appendices A and B.

\section{Models}

In spite of their simplicity, Go-like models \cite{Go} have been surprisingly
successful in capturing the protein folding mechanism
\cite{HoangJCP1,HoangJCP,Clementi,Baker}. 
In the version of Go-like model considered in the present study, each amino
acid is represented by a single bead located at position of C$_\alpha$ atom.
The potential energy of a given protein conformation is given by:
\begin{eqnarray}
E & = & \sum_{i=1}^{N-1} K_b (r_{i,i+1} - b)^2 +
\sum_{i=2}^{N-1} K_\theta (\theta_i - \theta_{i}^*)^2 + \nonumber \\
& &
+ \sum_{n=1,3} \sum_{i=2}^{N-2} K_\phi^{(n)} [1+\cos(n(\phi_i - \phi_{i}^*))] + 
\nonumber \\
& &
+ \sum_{i+3<j}
4\epsilon \left[ \left( \frac{\sigma_{ij}}{r_{ij}} \right)^{12} 
- \left( \frac{\sigma_{ij}}{r_{ij}} \right)^{6} \right] \Delta_{ij} +
\nonumber \\
& &
+ \sum_{i+3<j}
\epsilon \left( \frac{\sigma}{r_{ij}} \right)^{12} (1-\Delta_{ij}) \ ,
\label{eq:Go}
\end{eqnarray}
where $N$ is the number of beads; $r_{ij}$ is the distance between
beads $i$ and $j$; $\theta$ and $\phi$ are bond and dihedral 
angles associated with the residues; the star superscript corresponds to the
native state; $\Delta_{ij}$ is equal to 1 if there is a native contact between
$i$ and $j$ and equal to 0 otherwise. A native contact is defined if
the distance between two residues in the native state is less than 7.5 $\AA$.
The first three terms of Eq. (\ref{eq:Go}) correspond to the bonding
potentials, bond angle potentials and dihedral angle potentials, respectively.
The last two terms correspond to Lennard-Jones (LJ) potentials for native
contacts and repulsive potentials for non-native contacts. The choice of LJ
potential is such that the native distance between residues in a native
contact corresponds to the potential minimum, i.e. $\sigma_{ij}=2^{-1/6}
r_{ij}^*$. In simulations, a native contact is said to be formed if
$r_{ij} < 1.5 \sigma_{ij}$. $\epsilon$ is the depth of the LJ potential and sets
the energy unit. Following Ref. \cite{Clementi}, the parameters chosen for our
model are $b=3.8\AA$, $\sigma=5\AA$, $K_b=100\,\epsilon\AA^{-2}$,
$K_\theta=20\,\epsilon(\mathrm{rad})^{-2}$, $K_\phi^{(1)}=-\epsilon$, and
$K_\phi^{(3)}=-0.5\epsilon$.

\begin{figure}
\center
\includegraphics[width=3in]{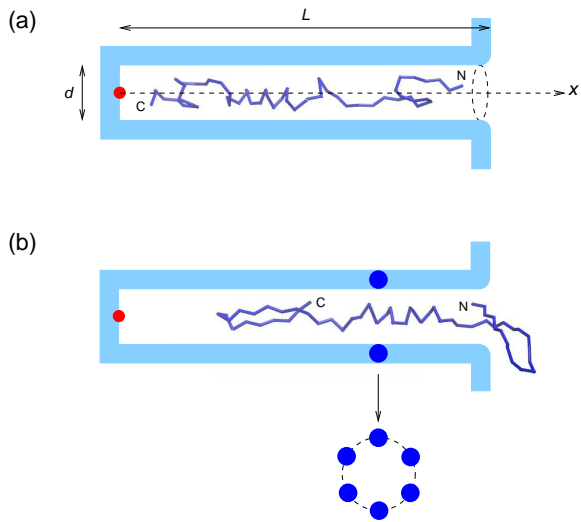}
\caption{(Color online) 
Schematic of models of ribosomal exit tunnel.
The tunnel is modelled as a hollow cylinder with one bottom close and
the other bottom open on a planar wall. The tunnel has
a length $L=100\AA$ and a diameter $d=15\AA$, and is centered along the $x$
axis. The tunnel's wall is either purely repulsive to amino acids (a), or to
contain
several attractive sites, occupied by residues (blue dots) of the same size as 
that of amino acid, equally spaced on a ring perpendicular to
the tunnel axis at $x=70 \AA$ (b). 
Nascent protein grows from the peptidyl transferase center (PTC) (red dot)
located at the axis origin and is released into the tunnel. The conformations
shown are snapshots taken from a MD simulation of protein GB1.
}
\label{fig:model}
\end{figure}

We model the exit tunnel as a half-close cylinder of length $L=100\AA$
and diameter $d=15\AA$ with an open bottom attached to a planar wall \cite{Thuy}
(Fig. \ref{fig:model}). The latter mimics the ribosome outer
surface and is repulsive to all amino acids.
The tunnel is assumed to be centered along the $x$ axis, whose origin is also
the position of the peptidyl transferase center
(PTC). The nascent chain is grown to its full length from the
PTC, and escapes from the tunnel through the open bottom.  
We will consider two models of the tunnel: a purely repulsive one and 
the one with attractive sites. In the first model (Fig. \ref{fig:model}a),
the interaction between the tunnel wall and an amino acid residue is given by
a repulsive truncated soft-core potential:
\begin{equation}
V_\text{wall} (r) = \left\{
\begin{array}{ll}
4\epsilon \left[ \left( \sigma/r \right)^{12} 
- \left( \sigma/r \right)^{6} \right] + \epsilon
& ,\quad r \leq 2^{1/6} \sigma \\
0 & , \quad r > 2^{1/6} \sigma
\end{array}
\right.
\label{eq:ur}
\end{equation}
where $\sigma=5\AA$ and $r$ is equal to the shortest (radial) distance from the
residue to the cylinder's wall plus 2.5 $\AA$. 
The added distance of 2.5 $\AA$ corresponds to the Van der Waals (VdW)
radius of a virtual `residue' embedded in the tunnel's wall, assumed to be
approximately of the same size as that of amino acid.
The same potential as given in Eq. (\ref{eq:ur}) is
used for the amino acid repulsion by the planar wall and the tunnel exit port.
In the second model (Fig. \ref{fig:model}b),
attractive sites are occupied by residues of VdW radius of 2.5 $\AA$
tangent to the tunnel wall's inner surface and equally spaced on a ring
perpendicular to the tunnel axis at $x=70 \AA$. The number of attractive
sites, $n_a$, is equal to either 4 or 6 in the present study. 
The residues at the attractive sites interact with amino acids via LJ
potential: 
\begin{equation}
V_\mathrm{LJ}(r) = 4\epsilon \left[ \left( \frac{\sigma}{r} \right)^{12} 
- \left( \frac{\sigma}{r} \right)^{6} \right] ,
\end{equation}
where $r$ is the distance between an attractive site and an amino acid.

The Langevin equations of motion for amino acids are integrated by using
Molecular Dynamics (MD) method with Verlet algorithm (Appendix A). 
In the simulations, the unit of time is $\tau=\sqrt{m\sigma^2/\epsilon}$,
where $m$ is the mass of amino acid (supposed to be uniform for all amino
acids). Temperature is given in units of $\epsilon/k_B$. The friction
coefficient $\zeta$ of an amino acid is given in units of $m\tau^{-1}$.
The free energy profiles of protein at the exit tunnel are calculated 
by using the combined umbrella sampling technique \cite{Torrie} with the
weighted histogram method \cite{Ferrenberg,Kumar} (see Appendix B).
In the free energy calculation,
a periodic boundary condition with a box size equal to
10$L$ is applied in directions perpendicular to the $x$ axis for $x>L$
(the space outside the tunnel).

The folding time and the escape time of protein are measured from the moment
the full length protein is released from the PTC. The protein is said to be
folded when all the native contacts are formed, and to be escaped when the
number of amino acids out side the tunnel, $N_\mathrm{out}$, is equal to the
total number of residues, $N$.  The median folding time, $t_\mathrm{fold}$, and
the median escape time, $t_\mathrm{esc}$, at a given temperature are determined
from multiple independent simulations (500-1000 trajectories).

\section{Results}

\begin{figure}
\center
\includegraphics[width=3.2in]{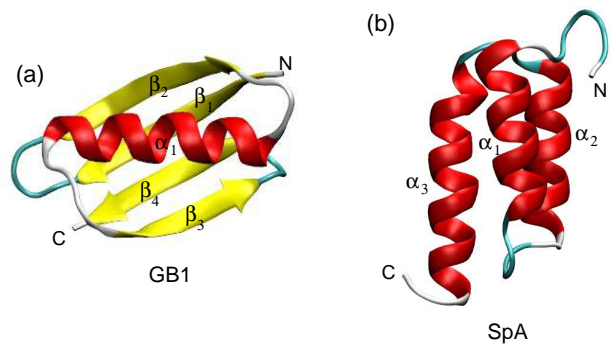}
\caption{(Color online) Ribbon presentation of the native state conformations
of the proteins considered in the present study: the B-1 domain of protein G
(GB1) (a) and Z domain of Staphylococcal protein A (SpA) (b). The 
corresponding PDB codes are 1pga and 2spz for GB1 and SpA, respectively.
The secondary structures are labeled as indicated.}
\label{fig:native}
\end{figure}

We will consider two small proteins: the B-1 domain of protein G (GB1)
of length $N=56$, and the Z domain of Staphylococcal protein A (SpA)
of length $N=58$. The native states of these two proteins are shown in
Fig.  \ref{fig:native}.  Our equilibrium
simulations for GB1 and SpA without the tunnel indicate that
they both have a two-state folding transition  as displayed by a sharp specific
heat peak and a bimodal free energy profile near the folding
transition temperature (Fig. S1 of supplemental material \cite{supp}).
The folding temperature, $T_f$, defined as the temperature of the maximum
of the specific heat peak, is found to be equal to 0.922 and 0.804 for GB1 and
SpA, respectively, in units of $\epsilon/k_B$. 

\begin{figure*}
\centering
\includegraphics[width=4.5in]{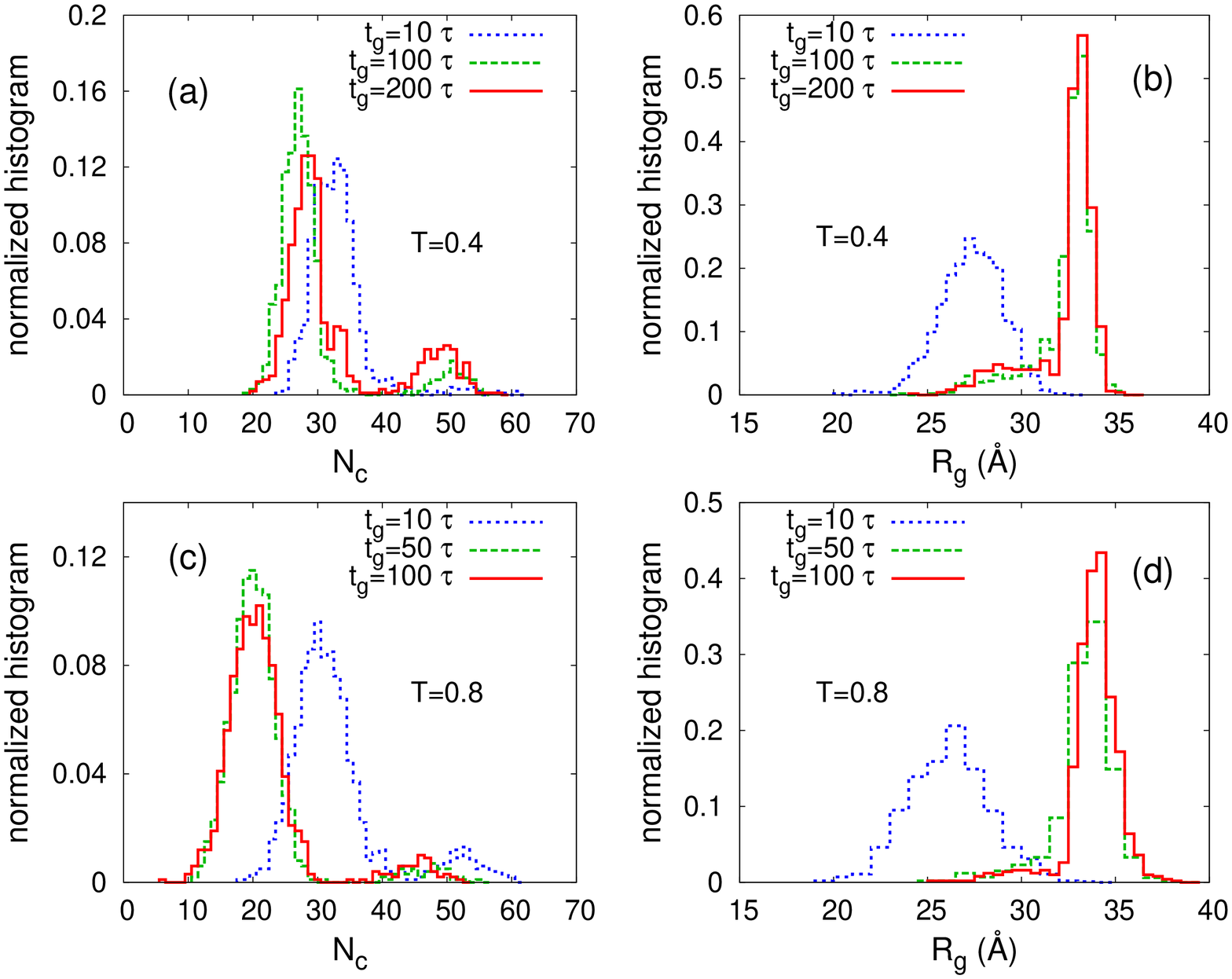}
\caption{
(Color online)
Histograms of conformations of the full length protein as result of
the chain growth process at the ribosomal tunnel, obtained at the moment of
complete translation, as functions of the number of native contacts, $N_c$, and
the radius of gyration, $R_g$, for GB1 at temperatures $T=0.4\
\epsilon/k_B$ (a,b) and $T=0.8\ \epsilon/k_B$ (c,d), as indicated. The
conformation ensembles are generated by 1000 independent growth simulations for
each temperature and for a given growth speed with $t_g$ defined as the time
needed for the chain elongation of one amino acid. Three different growth
speeds are shown for each temperature as indicated.
}
\label{fig:hisgrow}
\end{figure*}

\subsection{Folding and escape at a purely repulsive tunnel}

\begin{figure*}
\centering
\includegraphics[width=6.5in]{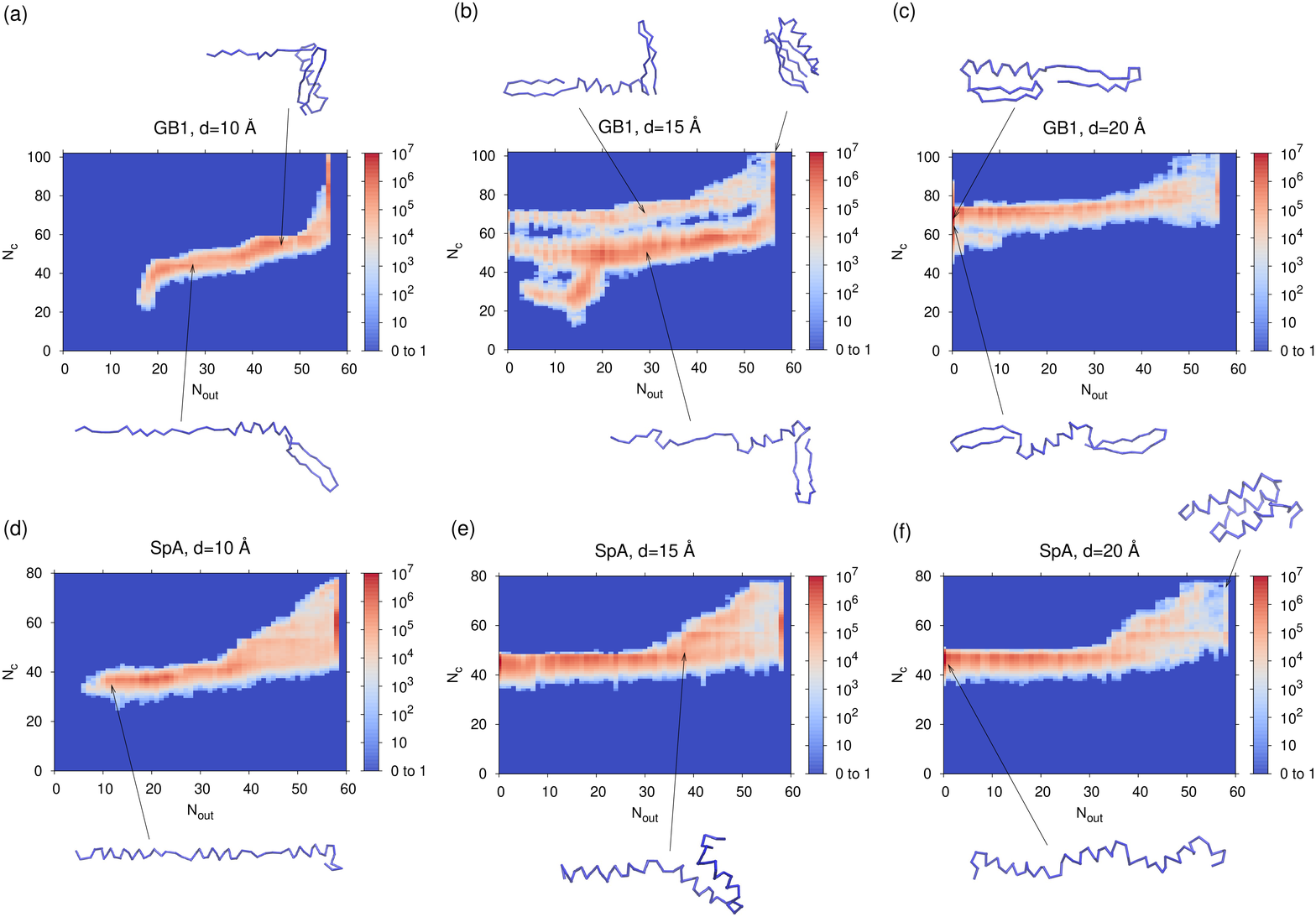}
\caption{(Color online)
Histograms of conformations during folding and escape of GB1 (a,b,c)
and SpA (d,e,f) as a function of the number of amino acids outside the
tunnel ($N_{out}$) and the number of native contacts formed ($N_c$) at a purely
repulsive exit tunnel at temperature $T=0.4 \epsilon/k_B$, for three different
tunnel diameters: $d=10\AA$ (a,d), $d=15\AA$ (b,e) and $d=20\AA$ (c,f).  The
histograms are obtained from 100 independent trajectories with the growth time
per amino acid of $t_g=100\tau$ during translation. The colors shown
correspond to the values of the histogram in a logarithmic scale.  Examples of
protein conformations obtained in the simulation are shown.
}
\label{fig:ghis}
\end{figure*}

We first study folding and escape of the proteins at a purely repulsive tunnel. 
The simulations start with a cotranslational protein folding inside the tunnel
with a steady growth of the polypeptide chain.  The translation speed is set by
$t_g$ -- the time needed for the chain elongation of one amino acid. Protein
synthesis in vivo typically takes times of several orders of magnitude longer
than the refolding time of a free protein in solution. Such a realistic
translation time cannot be reached by our simulations. Yet, we find that
moderate translation times up to about 10 times of the minimum refolding time
are sufficient to yield a protein conformation ensemble similar to that would
be obtained by a much slower translation. Fig. \ref{fig:hisgrow} shows
histograms of conformations as functions of the number of native contacts and
the radius of gyration obtained by multiple independent growth simulations for
protein GB1 at several growth speeds at two different temperatures lower than
$T_f$. For both temperatures, it is shown that $t_g=10\tau$ can be considered
as a fast translation as it yields conformations that contain more native 
contacts and are significantly more compact  than those obtained with larger
$t_g$. However, the histograms quickly converge as $t_g$ is increased to about
100$\tau$. At $T=0.4\epsilon/k_B$, the histograms are similar for $t_g=100\tau$
and $t_g=200\tau$ indicating that the effect of slow translation can be
captured already at $t_g=100\tau$. At $T=0.8\epsilon/k_B$, even $t_g=50\tau$ is
sufficient to yield slow translation conformations. Similar analysis for the
helical protein SpA (Fig. S2 of supplemental material \cite{supp}) indicates
that the growth with $t_g=50\tau$ can be considered as a slow translation. Note
that total growth time for $t_g=100\tau$ for GB1 is about 12 times
larger than its minimum refolding time.

After the translation is complete, the nascent protein is released from the PTC
and continues to fold while escaping from the exit tunnel. 
To elucidate how the folding and the escape take place at the exit tunnel, 
we calculated two-dimensional histograms of
conformations in multiple folding
trajectories as a function of the number native contacts, $N_c$, and the number
of amino acid outside the tunnel, $N_{out}$. These histograms at a low
temperature of $T=0.4\,\epsilon/k_B$ are shown in Fig. \ref{fig:ghis} for the
two proteins, GB1 and SpA, and for three different values of tunnel diameter
$d$ (only in this figure, $d$ is varied for comparison).
For all cases, it is shown that a large portion of native contacts is
formed when the protein is found completely or almost completely inside the
tunnel (small $N_{out}$). As $N_{out}$ increases, the number of native contacts
generally also increases indicating a concomitance between folding and escape.
The chain is completely folded only when all amino acids are outside the
tunnel. For GB1, the folding mechanism is
found to be strongly dependent on the tunnel diameter $d$. For $d=10\AA$, due
to the narrow confinement, only the $\alpha$-helix can form inside the
tunnel. For $d=15\AA$, both the $\alpha$-helix and the $\beta$-hairpin can form
inside the tunnel, but the latter is not always seen. For this diameter,
folding proceeds through two distinct pathways depending on whether
the C-terminal $\beta$-hairpin is formed inside the tunnel or not
(Fig.  \ref{fig:ghis}b), as also shown in Ref. \cite{Thuy} for a different
growth rate. For $d=20\AA$, 
a partial tertiary structure formed by an
$\alpha$-helix and $\beta$-hairpin is also found inside the tunnel (Fig.
\ref{fig:ghis}c). For SpA, the folding mechanism does not depend on
$d$. Folding of this protein
proceeds through a single pathway, with a formation of all the
$\alpha$-helices followed by successive formation of tertiary contacts as the
protein gradually escapes from the tunnel. The relatively simpler folding
mechanism obtained for SpA comparing to GB1 is due to the fact that SpA
has only $\alpha$-helices which can easily form inside the tunnel.
We find that as temperature is increased to $T=0.6\,\epsilon/k_B$, the
folding mechanisms of the two proteins remain the same but the histograms are
more spread indicating an increased stochasticity of the pathways.

Experimental studies have indicated that the
$\alpha$-helix and the $\beta$-hairpin can form inside the ribosomal exit
tunnel, while such a possibility is very limited for subdomain tertiary
structures \cite{Deutsch2005,Deutsch2009}.
Among the three diameters of the tunnel considered, $d=15\AA$ shows a folding
behavior the most consistent with these experimental evidences. 
$d=10\AA$ is too small for formation of the $\beta$-hairpin, while 
$d=20\AA$ is too large to inhibit the formation of tertiary structure
inside the tunnel. Thus, for the remaining of our study, we will consider only
the case of $d=15\AA$.

\begin{figure}
\center
\includegraphics[width=3in]{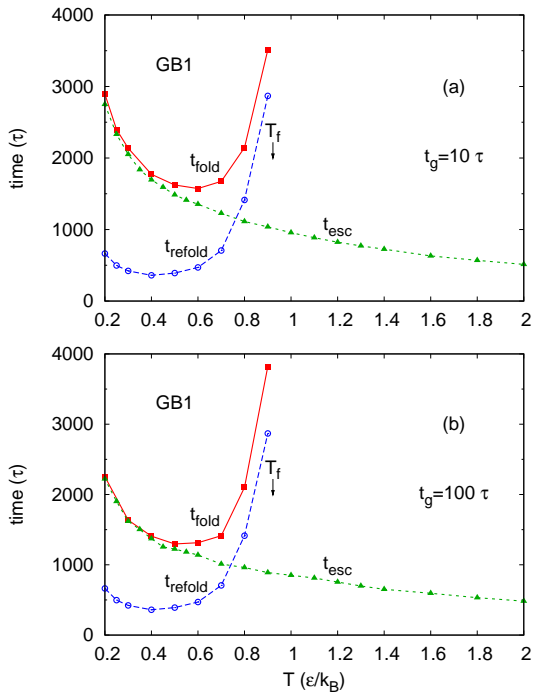}
\caption{(Color online) Temperature dependence of the median folding time at
a purely repulsive tunnel ($t_\text{fold}$), the median escape
time from the tunnel ($t_\text{esc}$), and the median refolding time of free
denatured protein ($t_\text{refold}$) for GB1.
The folding time and the escape time are measured from the
moment the full length nascent protein is released from the PTC,
and are obtained for two translation speeds: a fast
translation with the growth time per amino acid 
$t_g=10\tau$ (a) and a slow translation with $t_g=100\tau$ (b). 
The data at each temperature were obtained from 1000 independent folding
trajectories.
}
\label{fig:gfold}
\end{figure}

Fig. \ref{fig:gfold} (a and b) show the temperature dependence of the folding 
time and the escape time for protein GB1 at the exit tunnel for
two different growth speeds of the growth time per amino acid $t_g=10\tau$ and
100$\tau$. For comparison, we have calculated also the median refolding time,
$t_\mathrm{refold}$, 
from unfolding and refolding simulations without the tunnel. 
The unfolding was done at $T=2\ \epsilon/k_B$ yielding 
extended conformations ($R_g > 20 \AA$).
It is shown that $t_\mathrm{fold}$ is significantly longer than 
$t_\mathrm{refold}$, while both has the characteristic U-shaped dependence on
temperature.  For refolding without the tunnel, the temperature of the fast
folding, $T_\mathrm{min}$, is found to be about 0.4 $\epsilon/k_B$. 
For the folding at the tunnel, $T_{min}$ is about 0.6 $\epsilon/k_B$
and 0.5 $\epsilon/k_B$ for $t_g=10 \tau$ and 100$\tau$, respectively.
Thus, $T_{min}$ is shifted
towards higher temperature in the presence of the tunnel. This shift suggests
that the tunnel strongly affects the folding kinetics at low temperatures. The
effect of the tunnel can be seen more clearly by comparing the folding
time and the escape time. Fig. \ref{fig:gfold} also shows the temperature
dependence of the median escape time, $t_\mathrm{esc}$. At low temperatures,
the folding time is only slightly larger than the escape time, indicating
that folding is concomitant with the escape process. This concomitance
means that that most of the native contacts form during the escape process,
and the final stage of folding corresponds to an establishment of a few
contacts when the last residue gets out from the tunnel.
As temperature increases the difference between $t_\mathrm{fold}$ and
$t_\mathrm{esc}$ also increases, suggesting a deteriorating concomitance 
between folding and escape. Note that $t_\mathrm{fold}$ diverges as $T$
approaches the folding temperature $T_f$ while $t_\mathrm{esc}$ continues to
decrease monotonically with $T$.
Fig. \ref{fig:gfold} also shows that the slow translation with
$t_g=100\tau$ leads to shorter escape times and folding times than the fast
translation with $t_g=10\tau$, but the qualitative picture of their
temperature dependence is the same for the two cases.

\begin{figure}
\begin{center}
\includegraphics[width=3in]{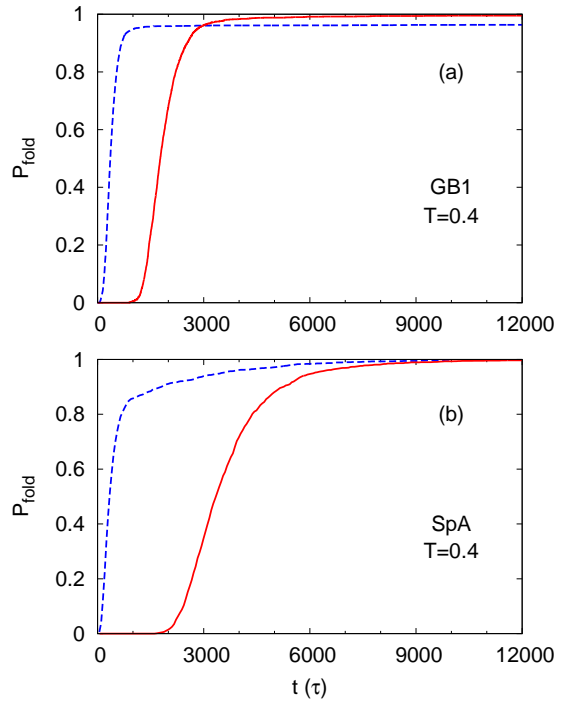}
\end{center}
\caption{(Color online)
Time dependence of the probability of successful folding, $P_\mathrm{fold}$, in
presence of a purely repulsive tunnel (solid line) and in refolding 
without the tunnel (dashed line) at $T = 0.4 \epsilon/k_B$
for GB1 (a) and SpA (b), respectively. For the folding at the
tunnel, the translation was carried out with $t_g=10\tau$ and time is measured
from the moment the full length protein is released from the peptidyl
transferase center.
}
\label{fig:pfold}
\end{figure}

Interestingly, the folding efficiency is enhanced in the presence of the
tunnel, at least for one of the proteins considered. Fig. \ref{fig:pfold}a
compares the probabilities of successful folding, $P_\mathrm{fold}$, as
functions of time for GB1 at $T=0.4\,\epsilon/k_B$ for two cases: folding at
the tunnel and refolding without the tunnel. In the former case,
$P_\mathrm{fold}$ increases more slowly with time but finally reaches a value
close to 1 at sufficiently long time. On the other hand, $P_\mathrm{fold}$ for
refolding without the tunnel can only reaches 0.96 at the time limit shown in
the figure ($10^4 \tau$). For SpA, $P_\mathrm{fold}$ reaches 100\% for both the
folding at the tunnel and refolding without the tunnel, after about $10^4 \tau$
(Fig. \ref{fig:pfold}b).

\begin{figure}
\centering
\includegraphics[width=3in]{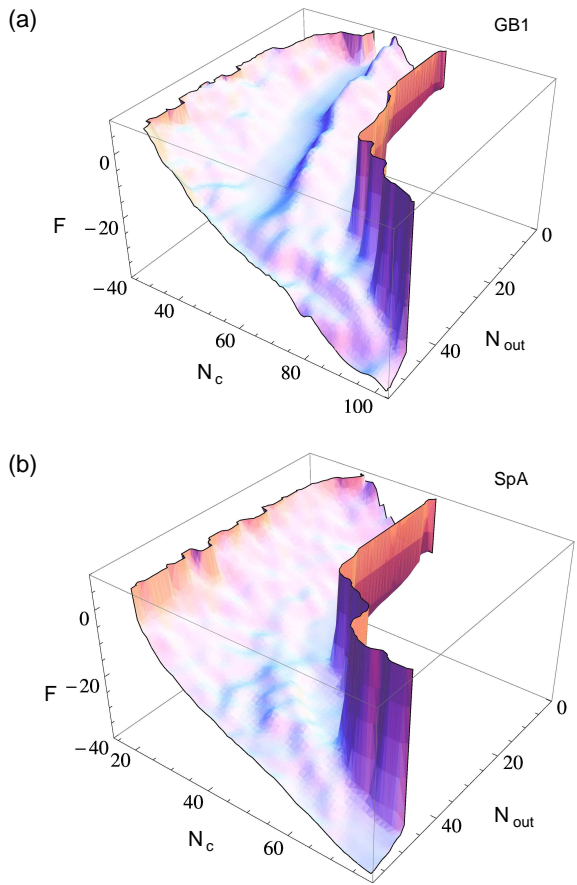}
\caption{(Color online)
Dependence of the free energy on the number of amino acids escaped from the
tunnel ($N_{out}$) and the number of native contacts ($N_c$) for 
GB1 (a) and SpA (b), respectively, at a purely
repulsive tunnel, at temperature $T=0.4\,\epsilon/k_B$. 
The range of $N_{out}$ shown is from 0 to $N-1$, where
$N$ is the number of amino acids of the protein.
The free energy was calculated based on umbrella
sampling simulations and the weighted histogram method (see Appendix B).
}
\label{fig:fsurf}
\end{figure}

The above dynamical analysis shows that at low temperatures, such as 
$T=0.4 \,\epsilon/k_B$, the folding and the escape processes are concomitant
with some variations due to the stochasticity of the two processes. 
Interestingly, our equilibrium free energy calculations based on umbrella
sampling of protein conformations along the exit tunnel also confirm this
scenario. Fig. \ref{fig:fsurf}a shows a free energy surface as function of
number of native contacts and then number of escaped amino acids for 
GB1 at $T=0.4\,\epsilon/k_B$. One can see that there are two channels 
separated by a significant barrier on the
free energy surface corresponding to the two dynamical pathways shown in Fig.
\ref{fig:ghis}b. Remarkably, the free energy landscape also shows that the
two pathways are downhill and without kinetic traps.  Fig. \ref{fig:fsurf}b
shows that for SpA, there is single pathway along the steepest descent
route in the free energy surface at $T=0.4\,\epsilon/k_B$, consistent
with the dynamical pathway shown in Fig. \ref{fig:ghis}e. This pathway is
also downhill and contains no kinetic traps.

\begin{figure}
\centering
\includegraphics[width=3.2in]{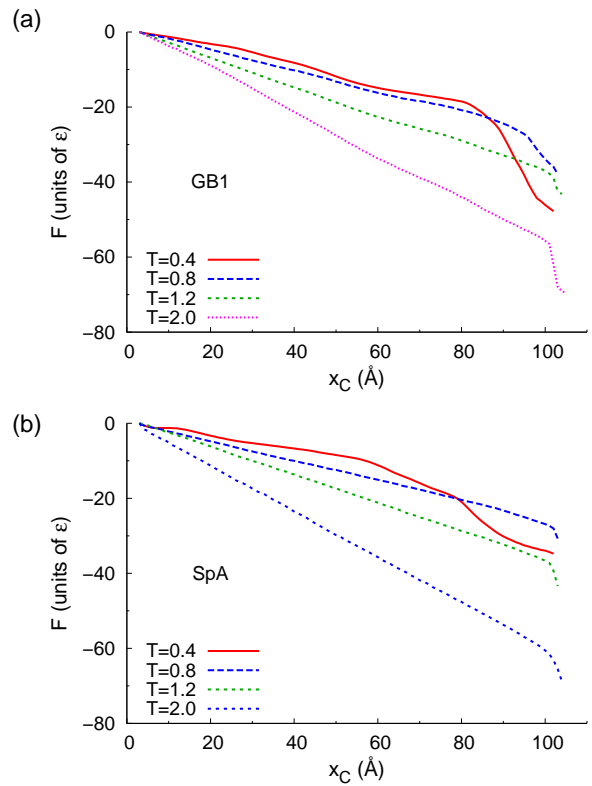}
\caption{(Color online)
Dependence of the free energy $F$ on the coordinate $x_C$ of the
$C$-terminus along the tunnel axis for the dominant pathway of
GB1 (a) and SpA (b) at a purely repulsive tunnel, at temperatures $T=0.4$,
0.8, 1.2 and 2.0 $\epsilon/k_B$ as indicated. 
}
\label{fig:ftemp}
\end{figure}

It is convenient to plot the free energy as a function of a single coordinate
along a given pathway. For GB1, our choices of such a coordinate are the $x$
coordinate of the C-terminus ($x_C$) for the pathway, in which the C-terminal
$\beta$-hairpin is not formed, and that of the 48th residue in the sequence
($x_{48}$) for the other pathway, in which the $\beta$-hairpin is formed. These
coordinates are chosen because, on average, the selected residues are the
latest ones to escape from the tunnel in the two pathways, respectively. For
SpA, $x_C$ is chosen as the coordinate for its only pathway. These coordinates
are restrained in our umbrella sampling simulations. Fig. \ref{fig:ftemp}a
shows the free energy profiles $F(x_C)$ for the dominant pathway of GB1,
obtained by umbrella sampling at four different temperatures: $T=0.4$,
0.8, 1.2 and 2.0 $\epsilon/k_B$. It is shown that for all temperatures
considered, $F$ is a smoothly decreasing function of $x_C$ with a slope
strongly dependent on temperature.  For temperatures higher than $T_f$, such as
$T=1.2$ and $T=2.0$, the dependence of $F$ on $x_C$ is almost linear over the
range of $x_C$ inside the tunnel.  This is due to the fact that the protein is
unfolded at these temperatures and the main contribution to the free energy
comes from the entropy of the escaped part of the chain.  On the other hand, as
temperature decreases below $T_f$, i.e. $T=0.8$ and $T=0.4$, the slope of the
$F(x_C)$ curve varies and depends on the position, suggesting an increasing
role played by the folding of the protein. At these temperatures, the potential
energy of the chain mainly contributes to the free energy.  Note that the free
energy difference between the fully escaped conformation ($x_C > L$) and the
initial conformation ($x_C \approx 0$) depends non-monotonically on
temperature. For the four temperatures shown, this free energy difference is
the smallest for $T=0.8$ because this temperature is the closest to $T_f$
(0.922), at which the effects of entropy and enthalpy are likely to
annihilate each other.
Similar scenario is seen for SpA (Fig. \ref{fig:ftemp}b).

\subsection{Diffusional description of the escape process}

The method of mapping complex dynamics into low dimensional diffusion models is
a powerful technique traditionally used in physics and chemistry, such as for
calculating the rate of chemical reaction \cite{Kramers}, for the analysis of
mean first passage time \cite{Zwanzig} and for identification of reaction
coordinates \cite{Edwards,Peters}. The free energy profiles shown in Fig.
\ref{fig:ftemp} are the potentials of mean force acting on the C-terminus of
nascent protein. Their monotonic behavior suggests that the escape process may
be described as a drift in an one-dimensional potential field.  Consider
diffusion of a particle in a potential field $U(x)$, which is
described by one-dimensional Smoluchowski equation given by
\cite{Nitzan}:
\begin{equation}
\frac{\partial}{\partial t}\, p(x,t|x_0,t_0) =  
\frac{\partial}{\partial x} D
 \left(\beta \frac{\partial U(x)}{\partial x} + \frac{\partial}{\partial x}
\right)\, p(x,t|x_0,t_0) ,
\label{eq:smolu}
\end{equation}
where $p(x,t|x_0,t_0)$ is a conditional probability density of finding the
particle at position $x$ and at time $t$, given that it was found previously
at position $x_0$ at time $t_0$; $\partial_t$ and $\partial_x$ are partial time
and space derivatives, respectively; $D$ is diffusion constant, assumed to be
position independent; and $\beta=(k_B T)^{-1}$ is the inverse temperature. For
conventional diffusion in viscous fluids, $D$ satisfies the
Einstein-Smoluchowski relation in the form of a fluctuation-dissipation theorem
\cite{Nitzan}:
\begin{equation}
D = \frac{k_B T}{\zeta} ,
\end{equation}
where $\zeta$ is the friction coefficient.
If $U(x)$ is a linear potential on the position:
\begin{equation}
U(x) = - k x ,
\end{equation}
then the Eq. (\ref{eq:smolu}) has a solution given by:
\begin{equation}
p(x,t)\equiv p(x,t|0,0)=\frac{D\beta k}{\sqrt{4\pi D t}} \exp\left[
-\frac{(x - D\beta k t)^2}{4Dt}
\right] ,
\label{eq:pxt}
\end{equation}
given that the initial condition is $p(x,0)=\delta(x)$.
From the above solution one gets the mean displacement of the particle: 
\begin{equation}
\langle x \rangle = (D\beta k) t \ ,
\end{equation}
with $D\beta k$ as diffusion speed.
The distribution of escape time of a nascent protein from the tunnel of length
$L$ can be given as $p(L,t)$, from which one obtains a mean escape time:
\begin{equation}
\mu_t = \langle t \rangle = \int_0^\infty t \, p(L,t)\, dt 
= \frac{2+\beta k L}{D (\beta k)^2} \ ,
\end{equation}
and the standard deviation:
\begin{equation}
\sigma_t \equiv (\langle t^2 \rangle - \langle t \rangle^2)^{\frac{1}{2}}
= \frac{\sqrt{8 + 2 \beta k L}}{D (\beta k)^2} \ .
\end{equation}
The ratio between $\sigma_t$ and $\mu_t$ is 
found to be independent of the diffusion constant:
\begin{equation}
\sigma_t = \frac{\sqrt{8 + 2 \beta k L}}{2 + \beta k L} 
\, \mu_t \ .
\label{eq:sigt}
\end{equation}
The median escape time, $t_\mathrm{esc}$, is defined by the following
equation:
\begin{equation}
\int_0^{t_\mathrm{esc}} p(L,t) \, dt = \frac{1}{2} \ .
\end{equation}
It can be easily shown that $t_\mathrm{esc} \sim D^{-1}$, while the dependence
of $t_\mathrm{esc}$ on $\beta k$ is not trivial.
In the simulations, due to the use of time cut-offs,
the median escape time can be calculated more conveniently and more accurately
than the mean escape time.

\begin{figure}
\centering
\includegraphics[width=3.3in]{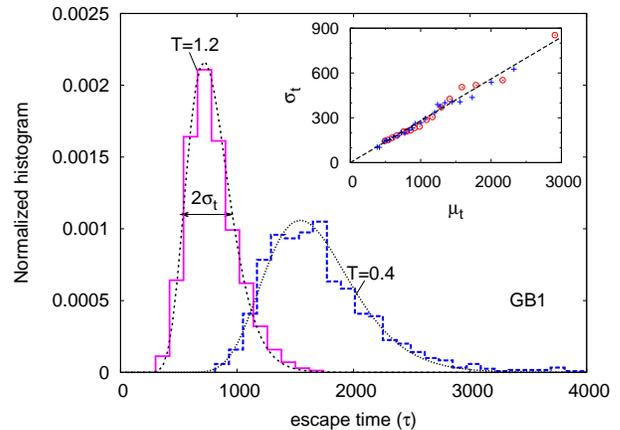}
\caption{(Color online)
Distribution of the escape time from a purely repulsive exit
tunnel for protein GB1. Normalized histograms 
shown for two temperatures, $T=1.2\,\epsilon/k_B$ (solid) and
$T=0.4\,\epsilon/k_B$ (dotted), are obtained
from simulations with the growth time per amino acid $t_g=10\tau$
during translation.  
The histograms are fitted (smooth lines) to the distribution
function given in Eq. (\ref{eq:pxt}). The inset shows that
the standard deviation, $\sigma_t$, of the escape time at various temperatures
increases linearly with its mean value, $\mu_t$, 
obtained with $t_g=10\tau$ (circles) and $t_g=100\tau$ (crosses),
respectively.
}
\label{fig:distrib}
\end{figure}

Fig. \ref{fig:distrib} shows that the distributions of escape time calculated
from our simulation data can be fitted well to the theoretical distribution
function given in Eq. (\ref{eq:pxt}). Interestingly, we have found that the
standard deviation of the simulation escape times, $\sigma_t$, depends linearly
on their mean value $\mu_t$ (Fig. \ref{fig:distrib}, inset). By fitting the data
to Eq. (\ref{eq:sigt}) with $L=100 \AA$, we find that $\beta k \approx 0.2537
\AA^{-1}$ and is approximately constant for various temperatures, 
and does not depend on the translation speed (Fig. \ref{fig:distrib}, inset).
Thus, $k$ is a linear function of $T$. Note that the linear form of $U(x)$ as
well as its dependence on temperature roughly agree with the shape of the
potentials of mean force obtained by umbrella sampling simulations shown in
Fig. \ref{fig:ftemp}.

\begin{figure}
\centering
\includegraphics[width=3.2in]{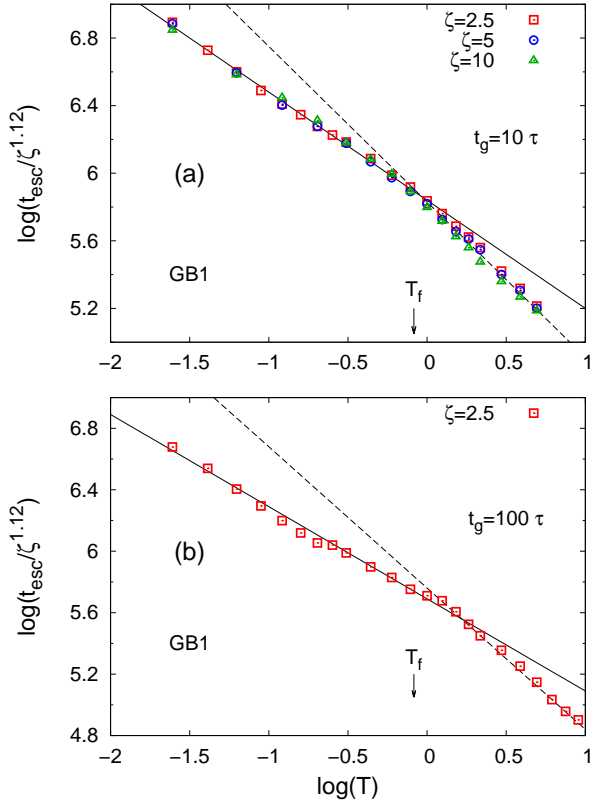}
\caption{(Color online)
(a) Log-log dependence of the median escape time on temperature for protein GB1
at a purely repulsive tunnel.
The plot shows a power law dependence, $t_\mathrm{esc} \sim \zeta^{1.12}
T^{-\alpha}$, with $\alpha = 0.64$ (solid line) for $T<T_f$, and
$\alpha = 0.92$ (dashed line) for $T>T_f$. Data points were obtained
from the simulations with the growth time per amino acid of
$t_g=10 \tau$ during translation, and correspond to different friction
coefficients of $\zeta = 2.5$, 5 and 10, in units of $m\tau^{-1}$,
as indicated. 
(b) Same as in (a) but with $t_g=100\tau$ and only the data
for $\zeta=2.5$ $m\tau^{-1}$ is shown. The power law fits shown are
obtained with $\alpha=0.6$ (solid line) and $\alpha=0.92$ (dashed line).
}
\label{fig:lgb1}
\end{figure}

\begin{figure}
\centering
\includegraphics[width=3.2in]{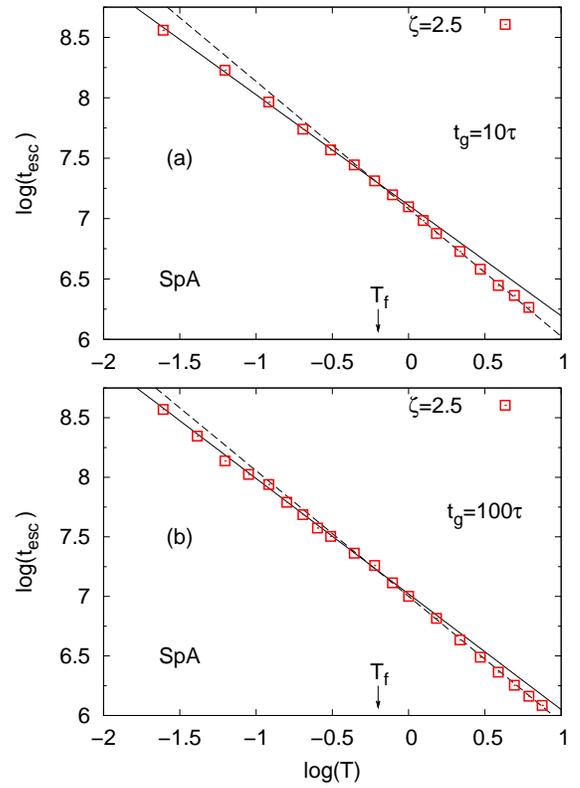}
\caption{(Color online) 
Same as Fig. \ref{fig:lgb1} but for SpA and only
the data for $\zeta=2.5 m\tau^{-1}$ is shown.
The power law fits in (a) are shown with the exponent $\alpha$ equal
to 0.915 and 1.005 for the low (solid line) and high (dashed line) temperature
regime, respectively. The corresponding
exponents of the fits in (b) are 0.97 and 1.005, respectively.
}
\label{fig:lspa}
\end{figure}

We found that the escape times obtained from the simulations at different
temperatures and different friction coefficient follow a common power law
dependence given by:
\begin{equation}
 t_\mathrm{esc} \sim \frac{\zeta^{1.12}}{T^\alpha} ,
\label{eq:tesc}
\end{equation} 
where the exponent $\alpha$ depends on the protein and on the temperature
regime, which is either lower or higher than the folding temperature $T_f$.
Fig. \ref{fig:lgb1}a shows this power law dependence for GB1 
with $\alpha \approx 0.64$ for $T<T_f$, and $\alpha \approx 0.92$ for $T>T_f$
for fast translation with $t_g=10\tau$. The collapse of the data for different values of friction coefficient $\zeta$ 
suggests that the simulations are done in the overdamped limit \cite{Klimov}.
The smaller exponent $\alpha$ for temperatures below the folding temperature
suggests that diffusion in this regime is facilitated by the folding of the
protein. If the protein was unfolded, the escape time in these
temperature range would be longer, as can be seen by extrapolation of the power
law for the higher temperature regime to the lower temperature regime (dashed
line in Fig. \ref{fig:lgb1}). 
Similar results are found for slow
translation with $t_g=100\tau$ (Fig. \ref{fig:lgb1}b), in which
the exponent $\alpha$ in the high temperature regime is the same as in the case
of fast translation ($\alpha \approx 0.92$) while it is
somewhat smaller in the low temperature regime ($\alpha \approx 0.6$).
Fig. \ref{fig:lspa} confirms the power law
dependence for protein SpA with the exponent $\alpha$ equal to 0.915 and 1.005
for the low and high temperature regimes, respectively, in the case of
fast translation.  For slow translation, $\alpha$ is equal to 0.97
and 1.005, respectively. Again, 
the diffusion of SpA at low temperatures is also facilitated by the folding
of this protein, but not as much as for GB1. The larger exponents found for SpA
than for GB1 suggest that diffusion of helical proteins is slower than
$\beta$-sheet containing proteins.  This reflects the fact that $\beta$-sheets,
particularly hairpins, are much more difficult to form inside the tunnel than
$\alpha$-helices.  Therefore, $\beta$-sheets are much more favorable when found
outside the tunnel, leading to an increased escape rate in comparison to
$\alpha$-helices.

The data on escape times reveal the dependence of diffusion constant
$D$ on temperature and friction coefficient. Because the diffusion
speed should be proportional to the inverse of escape time, one obtains:
\begin{equation}
D\beta k \sim \frac{T^\alpha}{\zeta^{1.12}} \ .
\label{eq:dbk}
\end{equation}
Since $\beta k$ is approximately constant in the present case of 
a purely repulsive tunnel, the behavior of $D$ is expressed as in
the r.h.s. of Eq. (\ref{eq:dbk}). The variation of the exponent $\alpha$ 
from the Einstein's value of one underscores for the change of protein
conformation during the escape process. Note that the dependence of $D$ on
friction coefficient $\zeta$ is slightly stronger than that of the
Einstein-Smoluchowski
relation.  This enhancement in the exponent of $\zeta$ can be understood as due
to additional friction between nascent protein and the tunnel wall.

\subsection{Effects of attractive sites}

We proceed now to consider the tunnel with attractive sites. Remind that our
model has several attractive sites located on the tunnel's wall at 30$\AA$
inwards from the tunnel exit (Fig.  \ref{fig:model}b). The positions of the
attractive sites were chosen such that they roughly coincide with the position
of the free energy minimum calculated for amino acids along the tunnel as shown
in Fig. 3 of Ref.  \cite{Pande2008}. In our model, these attractive sites
attract all amino acids,
therefore will certainly slow down the escape of nascent protein from the
tunnel. The question we ask is how the folding of protein at the exit tunnel
and how the diffusion characteristics of the escape process are affected by
attractive sites.  For simplicity, we consider only
the case of fast translation with $t_g=10\tau$ for the proteins at the
tunnel with attractive sites, with an assessment that the results of slow
translation are qualitatively similar.

\begin{figure}
\centering
\includegraphics[width=3in]{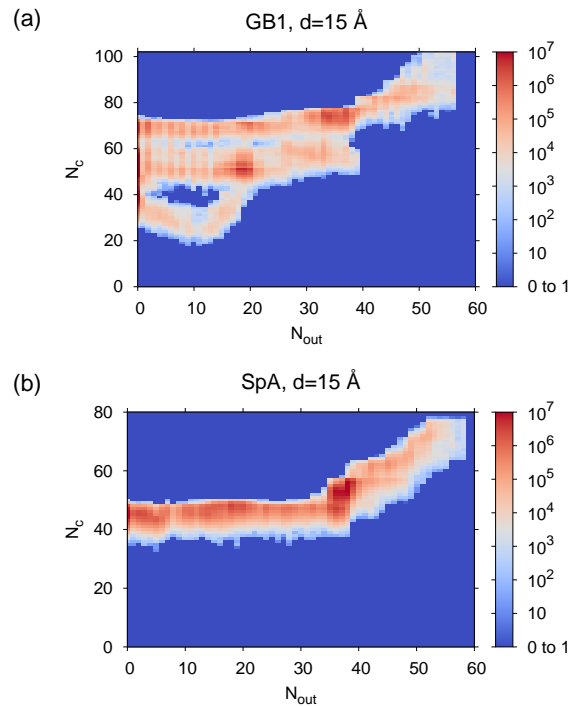}
\caption{(Color online)
Histogram of conformations obtained during folding and escape of 
protein GB1 (a) and SpA (b) at
the tunnel with 6 attractive sites. The histograms were computed from 100
independent folding trajectories at temperature $T=0.4\,\epsilon/k_B$ 
after a translation with $t_g=10\tau$.
}
\label{fig:ghis6}
\end{figure}

Our simulations show that the attractive sites can change the folding pathway
of GB1 while they have no such effect on SpA. Fig. \ref{fig:ghis6}a
shows the histogram of conformations obtained in multiple folding trajectories
at the tunnel with 6 attractive sites for GB1 at $T=0.4\,\epsilon/k_B$. In
contrast to the two distinct pathways found for the case of a purely repulsive
tunnel (Fig.  \ref{fig:ghis}b), one can see that in Fig. \ref{fig:ghis6}a,
there is only one pathway that exits from the tunnel. This pathway corresponds
to conformations in which
the C-terminal $\beta$-hairpin of GB1 is formed within the tunnel.  
For the tunnel with 4 attractive sites (Fig. S3 of supplemental material
\cite{supp}), the second pathway, in which the C-terminal $\beta$-hairpin is
not formed in the tunnel,
still exists but is populated by fewer trajectories than in the case of a
purely repulsive tunnel. Thus, the attractive sites promote formation of
$\beta$-sheet inside the tunnel. In the case of SpA, 
the $\alpha$-helices are easily formed inside the tunnel even without
the attractive sites, therefore the latter induces no change on the folding
pathway (Fig. \ref{fig:ghis6}b). However, since the attractive sites keep the
protein longer with the tunnel, they give more chance for the helices to form.
Note that for both proteins, the histograms for the tunnel with
attractive sites are less spread near the full escape ($N_\mathrm{out} >
40$) than for a purely repulsive tunnel indicating that the concomitance
between folding and escape is increased in presence of the attractive sites.
So the overall effects of attractive sites are an enhancement of secondary
structure formation inside the tunnel and an improved vectorial folding.

\begin{figure}
\centering
\includegraphics[width=3.2in]{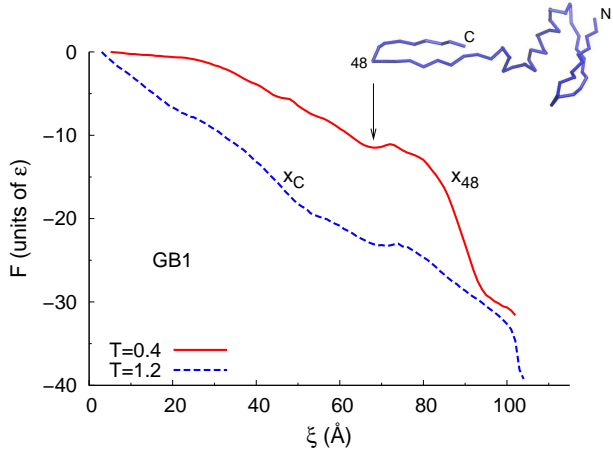}
\caption{(Color online)
Free energy $F$ as function of the restrained coordinate $\xi$,
being the $x$ coordinate the 48th amino acid in the sequence ($x_{48}$)
or that of the C-terminus ($x_C$), 
for protein GB1 at the exit tunnel with 6 attractive sites at temperatures
$T=0.4\,\epsilon/k_B$ (solid) and $T=1.2\,\epsilon/k_B$ (dashed), respectively.
The protein conformation shown corresponds to a local minimum in the free
energy as indicated.
}
\label{fig:fxcgb1}
\end{figure}

The escape times are found to be increased in the presence of the attractive
sites. The slowdown of escape due to the attractive sites can be understood by
looking at the free energy profiles shown in Fig. \ref{fig:fxcgb1} for GB1. A
small free energy barrier is seen on the escape route of the
upper pathway (in which the C-terminal $\beta$-hairpin is formed) for
$T=0.4\epsilon/k_B$ indicating that the attractive sites hamper the escape
process at the final stage. This hampering would allow a large portion
of tertiary contacts to form before the final release of the protein from the
exit tunnel. This barrier almost disappears for temperature higher than $T_f$
(the case of $T=1.2\,\epsilon/k_B$ in Fig.  \ref{fig:fxcgb1}) as a result of a
weaker interaction between the attractive sites and the unfolded protein and an
increased role of entropy. 

\begin{figure}
\centering
\includegraphics[width=3.2in]{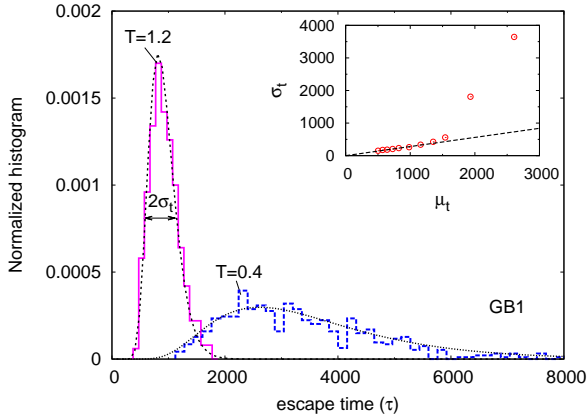}
\caption{(Color online)
Histograms of escape time for protein GB1 at the tunnel with 4 attractive
sites at $T=1.2 \epsilon/k_B$ (solid line) and $T=0.4\epsilon/k_B$ (dotted
line). The histograms are fitted (smooth lines) to the distribution function
given in Eq. (\ref{eq:pxt}). The standard deviation of the escape times
is plotted against their mean value for various temperatures (inset).
The translation was carried out with $t_g=10\tau$.
}
\label{fig:gtes4h}
\end{figure}

\begin{figure}
\centering
\includegraphics[width=3.2in]{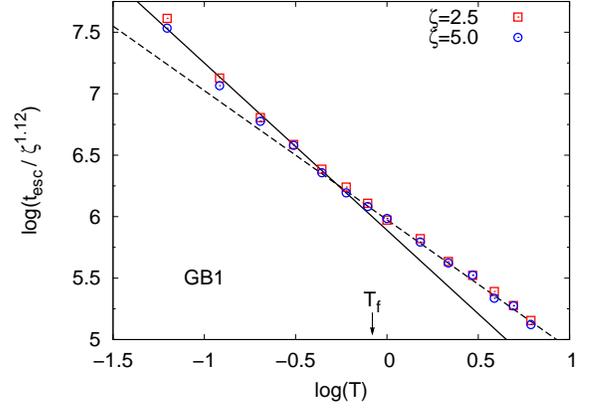}
\caption{(Color online)
Log-log dependence of escape time on temperature for protein GB1 
at the tunnel with 4 attractive sites. The data, obtained from simulations
with two different friction coefficients, $\zeta=2.5$ (squares) and 
$\zeta=5$ (circles), are fitted to the power law (Eq. (\ref{eq:tesc})) with
exponent $\alpha=1.33$ (solid line) and $\alpha=1.02$ (dashed line), for two
temperature ranges, lower and higher the the folding temperature $T_f$,
respectively. 
The translation was carried out with $t_g=10\tau$.
}
\label{fig:log4h}
\end{figure}

Fig. \ref{fig:gtes4h} shows that the escape process can be still 
relatively well described by the one-dimensional diffusion model for 
the tunnel with attractive sites still. It is shown that the
distribution of escape time can be fitted to the theoretical distribution
as given in Eq. (\ref{eq:pxt}). The linear dependence of the standard deviation
of the distribution on the mean escape time, however, is true only at
high temperatures or equivalently for low $\mu_t$ (Fig. \ref{fig:gtes4h},
inset). At low temperatures, this dependence becomes non-linear with the
standard deviation increased much faster than the mean. As a result, the
distribution at low temperatures,
such as at $T=0.4\,\epsilon/k_B$ as shown in the figure, is much more expanded
than it would be, based on a linear projection from high temperature
distributions.  Effectively, the parameter $\beta k$ of the effective potential
field $U(x)$
for the diffusion is not constant for the tunnel with attractive sites.

Fig. \ref{fig:log4h} shows that the dependence of the median escape time on
temperature and the friction coefficient still follows the power law given in
Eq. (\ref{eq:tesc}), but with larger exponents $\alpha$ than in the case 
of a purely repulsive tunnel
for both low and high temperature regimes. Note that 
the exponent $\alpha$ for the low temperature regime ($T < T_f$) is now
larger than that for the high temperature regime ($T > T_f$) in the case of a
tunnel with attractive sites. The higher exponent at low $T$ is also a
manifestation of the negative effect of attractive sites on diffusion. 
For the repulsive tunnel, we have shown that the folding of
nascent proteins induces a faster escape from the tunnel.
Here, the escape is slower due to the competition from the attractive sites 
that prevails the effect of folding.

The increase of escape time as well as the expansion of its distribution at low
temperatures can be understood as due to an increasing impact of the free
energy barrier shown in Fig. \ref{fig:fxcgb1}. At very low temperatures,
it is expected that the escape rate is limited by an activated diffusion 
through this barrier. In that case, the escape time should grow exponentially
with the inverse temperature \cite{Kramers,Klimov}. In the temperature range
studied, however, we have not yet observed this exponential slow-down.

\section{Discussion}

Protein cotranslational folding has been mainly discussed in the literature as
attributed to vectorial folding of nascent proteins resulted from the growth of
the polypeptide chain during translation \cite{Baldwin}. Here, we show that
vectorial folding is also induced by the ribosomal exit tunnel after the
translation is complete, as the nascent protein gradually escapes from the
tunnel.  Since simple secondary structures can form inside the tunnel (though
with some difficulty for the $\beta$-hairpin), the vectorial folding induced by
the tunnel mainly applies to the formation of the tertiary structures.  On the
other hand, for small single-domain proteins, such as the ones considered in
the present study, the vectorial folding due to translation would apply only to
the secondary structures, because a full length protein would still fit inside
the tunnel when the translation is complete.  For large multi-domain proteins,
both the peptide elongation and the escape from the tunnel are expected to give
rise to vectorial folding of protein tertiary structures. Our results are in
agreement with a previous simulation study \cite{Elcock} of on-ribosome folding
with a more realistic model of the exit tunnel, which showed that single-domain
proteins fold post-translationally. The different effects of cotranslational
folding on single- and multi-domain proteins have been also shown in Ref.
\cite{Elcock}.

Post-translational vectorial folding at the tunnel would not happen if the
escape time is much shorter than the relaxation time of the protein. In such a
case, the protein would be still largely unfolded when the escape is complete.
Folding would be concomitant with the escape process only if the latter is
sufficiently slow, so that the chain has enough time to relax during the
escape. Our simulations showed that, for temperatures about or lower than
$T_\mathrm{min}$ -- the temperature of the fastest folding in presence of the
tunnel, the escape time is significantly longer than the refolding time, thus,
allows for concomitant folding and escape at the tunnel. The fact that
vectorial folding induced by the tunnel happens at low temperatures is
beneficial for nascent proteins because it would help the protein to avoid
kinetic traps \cite{Bryngelson} which are easy to be encountered at low
temperatures. The escape time is expected to depend on the tunnel
size and shape. In fact, our simulations indicate that the escape time
increases with the tunnel diameter $d$ and length $L$ (detailed analysis will
be given elsewhere). It is an interesting question of whether the ribosomal
tunnel's detailed size and shape have been selected by Nature for a facilitated
vectorial folding of nascent proteins at the tunnel by mediating their escape
time.

Denatured proteins typically refold to their native states through multiple
pathways in a rugged energy landscape \cite{Bryngelson,Dill,HoangJCP}.  Here,
we show that folding at the tunnel proceeds along a single or at most a few
pathways in a much reduced conformational space. This dramatical reduction in
the number of conformations and the number of pathways is a key factor in
helping nascent proteins to avoid kinetic traps. Our free energy calculations
also showed explicitly that the pathways at the tunnel are downhill without
kinetic traps.  Remarkably, our simulations also showed that, at least for
protein GB1, the folding efficiency in presence of the tunnel is improved in
comparison to that without the tunnel. This result is in agreement with recent
experiments which indicated that the ribosome promotes correct folding of
nascent proteins \cite{Bustamante,Ugrinov}.

Our finding that the escape process can be mapped to an one-dimensional
diffusion model in an effective linear potential could be verified
experimentally and may be useful for studying the interactions between nascent
protein and the exit tunnel. We have shown that, in the case of the tunnel
without attractive sites, folding facilitates the escape process, as expressed
by a lower exponent $\alpha$ for the diffusion at temperatures lower than the
folding temperature than for that at higher temperatures.  In contrast, for the
tunnel with attractive sites, the exponent $\alpha$ is higher in the low
temperature range.  The attractive sites also lead to a non-linear dependence
between the standard deviation and the mean value of the escape time at low
temperatures. Thus, interactions of nascent proteins with the tunnel strongly
affect the dynamics of their escape from the tunnel. An analysis of escape
rates, that would be obtained by experiments, may reveal about these
interactions.

While the geometry of the ribosomal exit tunnel has been precisely determined
\cite{Voss}, its detailed interactions with nascent proteins are still under
ongoing research. In a recent study, Pande and coworkers
\cite{Pande2008} calculated free energy profiles for single amino acids in the
exit tunnel and found that a large free energy barrier (up to about 7 $k_B T$)
near the exit port is commonly observed for most of the amino acids, despite
the differences in their side-chain chemico-physical properties. They suggested
that this exit barrier is relevant to a gate-latch mechanism involving
ribosomal protein L39 of the exit tunnel.  Our result for the tunnel with
attractive sites also yields a free energy barrier for the nascent proteins.
Our analysis suggests that the role of this barrier is to slow down the escape
process of nascent proteins, thus strengthens the condition for vectorial
folding at the exit tunnel. The net effect of this barrier is an enhanced
foldability of the released protein and minimizing the chance of aggregation.

\section{Conclusion}

Folding of nascent proteins at ribosomal exit tunnel has been studied
in simple models.  We have shown that the ribosomal exit tunnel has several
strong effects on the dynamics of nascent proteins. First, it induces vectorial
folding with improved folding efficiency by allowing the nascent chain to
gradually escape from the tunnel. This vectorial folding is concomitant with
the escape process and
happens only at a low temperature range favorable for folding. Second, folding
at the tunnel follows a single or a few pathways allowing the protein to avoid
kinetic traps.  Third, the escape process can be characterized as 
one-dimensional diffusion of a particle in an effective linear potential field
which depends on the condition for folding, such as temperature, and the
interaction of nascent protein with the exit tunnel. It was shown that folding
speeds up the diffusion, whereas attractive interactions between the nascent
protein and the tunnel wall slow it down. $\alpha$-helical protein also tends
to escape more slowly than $\beta$-sheet containing protein. We have also shown
that a slow escape from the tunnel is better for vectorial folding,
and consequently improves folding efficiency. Thus, the ribosome may impose
large free energy barrier along the escape route of nascent protein at the exit
tunnel, for the purpose of enhancing protein foldability and preventing early
release of unfolded nascent proteins.

\section*{ACKNOWLEDGEMENT}

This research was supported by Vietnam National Foundation for Science and
Technology Development (NAFOSTED) under Grant No. 103.01-2013.16.
The simulations were carried out using the High Performance
Computing facilities at Center for Informatics and Computing of VAST.

\section*{Appendix A: Langevin dynamics with Verlet algorithm}

Simulations are carried out using molecular dynamics method with 
Langevin equations of motion. For a given bead,
the Langevin equation in one dimension is given by:
\begin{equation}
m\ddot{x}(t) = f(t) - \zeta \dot x(t) + \Gamma(t) \ ,
\end{equation}
where $x$ is the position of a bead, $m$ is its mass (assumed to
be equal for all amino acids), $f$ is the molecular force acting
on the bead and $\zeta$ is the
friction coefficient. $\Gamma$ is a random force induced by the solvent
in form of a white noise with zero mean, $\langle \Gamma \rangle = 0$, and the
time auto-correlation function satisfying the fluctuation-dissipation theorem:
\begin{equation}
\label{eq:9}
\langle \Gamma(t) \Gamma(t+t')\rangle = 2\zeta k_B T \delta(t')\ ,
\end{equation}
where $T$ is absolute temperature. 

A Verlet algorithm is developed to numerically integrate the Langevin
equations of motion. The position at time $t+\Delta t$ is given by:
\begin{eqnarray}
\label{eq:7}
x(t + \Delta t) & = & 
\left\{2 x(t) 
- \left(1 - \frac{\Delta t}{2m}\zeta \right) x(t - \Delta t) 
\ + \right. \nonumber \\
 &  &
\left. + \ \frac{\Delta t^2}{m}\left[ f(t) + \Gamma(t) \right] \right\}
 \left(1+\frac{\Delta t}{2m}\zeta\right)^{-1},  
%\label{eq:8}
%\dot x(t) & = & \frac{x(t+\Delta t)-x(t-\Delta t)}{2\Delta t} \ .
\end{eqnarray}
where $\Delta t$ is the integration time step.
Because
$\int_{-\infty}^{+\infty} \delta(t) dt = 1$, Eq.
(\ref{eq:9}) is given in a discrete form given by:
\begin{equation}
\langle \Gamma(t) \Gamma(t+ n\Delta t)\rangle =
\frac{2\zeta k_B T}{\Delta t} \delta_{0,n}\ ,
\end{equation}
where $\delta_{ij}$ is Kronecker's delta function and
$n$ is an integer.
Temperature is given in units of $\epsilon/k_B$,
whereas time is measured in units of $\tau=\sqrt{m\sigma^2/\epsilon}$.
The integration step is taken to be $\Delta t = 0.002 \tau$.
Typical simulations are done with friction coefficient 
$\zeta=2.5 \, m\tau^{-1}$.

\section*{Appendix B: Umbrella sampling}

To sample protein configurations that are partially located inside the
tunnel, umbrella sampling technique \cite{Torrie} is employed. One carries
out $R$ simulations at a set of temperatures $\{T_j, j=1,\ldots,R \}$ with the
following restraint potentials acting on a given residue of
the protein:
\begin{equation}
V_j(\xi) = \frac{1}{2} \lambda_j (\xi - \xi_{0,j})^2 ,
\quad
(j=1,2,\ldots,R)
\end{equation}
where $\xi$ is the $x$-coordinate of the residue, $\lambda_j$ are restraint
parameters chosen to be in the range of 
$[0.1,1.0]\ \epsilon\AA^{-2}$ (stronger restraint is applied near the exit
port), and $\xi_{0,j}$ are another restraint parameters of values chosen
between $0$ and $L$. For GB1, the restrained residue is chosen to
be either the last residue (the C-terminus) or the 48th amino acid in the
sequence for the efficient sampling along the two pathways (see Fig.
\ref{fig:ghis}b), respectively. For SpA, only the C-terminus
is being restrained.

The free energy (or potential of mean force) can be calculated using the
weighted histogram method with the WHAM
equations given by \cite{Ferrenberg,Kumar}:
\begin{equation}
P_{\beta,j}(E,\xi) = \frac{\sum_{k=1}^R N_k(E,\xi) \exp[-\beta(E+V_j(\xi))]}
{\sum_{m=1}^R n_m \exp[f_m -\beta_m(E+V_m(\xi))]} ,
\label{eq:wham1}
\end{equation}
and
\begin{equation}
\exp(-f_j) = \sum_E \sum_\xi P_{\beta_j,j} (E,\xi) ,
\label{eq:wham2}
\end{equation}
where $E$ is the energy of protein conformation without counting restraint
potential, $N_k(E,\xi)$ is the histogram collected in simulation $k$, 
$n_m$ is the total number of snapshots in simulation $m$, and 
$\beta=(k_B T)^{-1}$ is the inverse temperature. $f_j$ are calculated
from Eqs. (\ref{eq:wham1}, \ref{eq:wham2}) self-consistently.
Assume that $j=0$ corresponds to $V_0(\xi)=0$, the free energy of an
unrestrained protein as function of $\xi$ at a given temperature $T$ 
is given by:
\begin{equation}
F(\xi) = -k_B T \log \sum_E P_{\beta,0}(E,\xi) .
\end{equation}
The dependence of $F$ on other coordinates, such as the
number native contacts, $N_c$, and the number of escaped amino acids from the
tunnel, $N_{out}$, can be calculated from an weighted probability based on the
histograms collected along these coordinates in addition to $E$
and $\xi$:
\begin{equation}
F(N_c,N_{out}) = -k_B T \log \sum_\xi \sum_E P_{\beta,0}(N_c,N_{out},E,\xi) .
\end{equation}

\clearpage

\newpage

\setcounter{equation}{0}
\renewcommand\theequation{S\arabic{equation}}

\setcounter{figure}{0}
\renewcommand\thefigure{S\arabic{figure}}

\onecolumngrid

\begin{center}
{\large\bf Supplemental material for: Folding and escape of nascent proteins
at ribosomal exit tunnel}
\end{center}

\centerline{Bui Phuong Thuy and Trinh Xuan Hoang}

\vspace{30pt}

\begin{figure}[!ht]
\centering
\includegraphics[width=6.in]{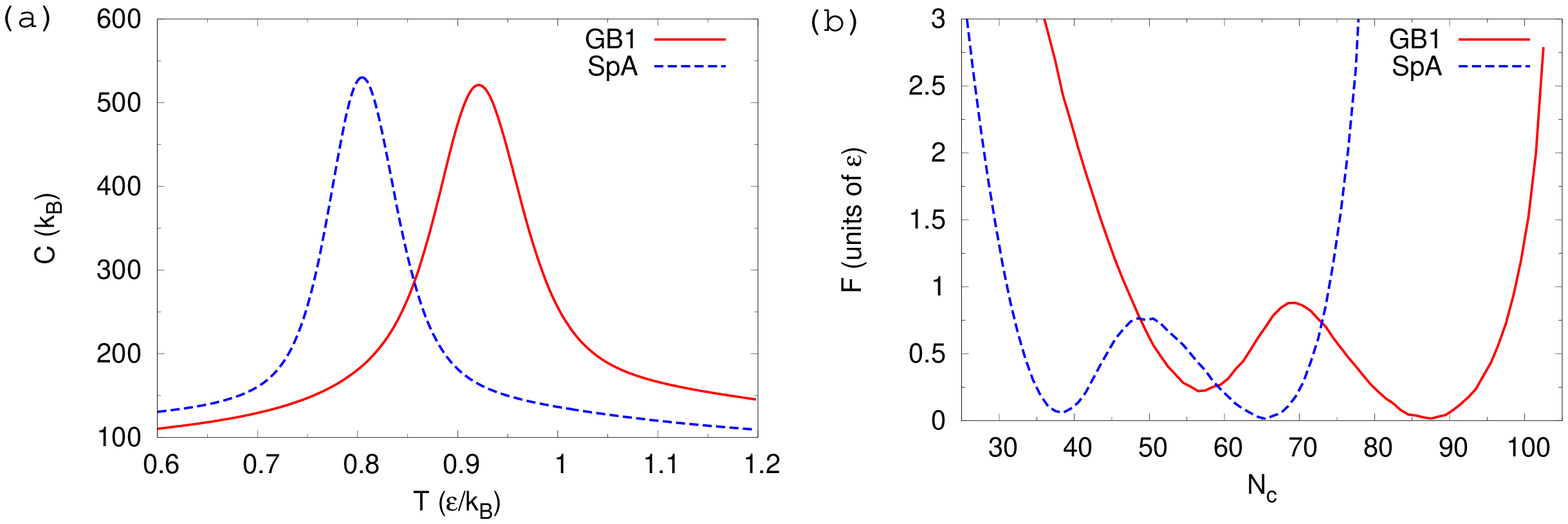}
\caption{(a) Temperature dependence of the specific heat, $C$, for proteins
GB1 (solid line) and SpA (dashed line). The specific heat was calculated
from the data of parallel tempering simulations of the proteins 
without the tunnel by using of the weighted histogram method.
The folding temperature, $T_f$, defined as the temperature of the specific
heat peak, is found to be equal to 0.922 and 0.804 for GB1 and SpA,
respectively, in units of $\epsilon/k_B$.
The relatively larger $T_f$ found for protein GB1 is due to the fact that,
in the present model, the number of native contacts in this protein is higher
than in SpA (102 vs. 78).
(b) Dependence of the effective free energy, $F$, on the number of native
contacts, $N_c$, at $T=T_f$ for the two proteins considered.
}
\end{figure}

\begin{figure}[!ht]
\centering
\includegraphics[width=4.5in]{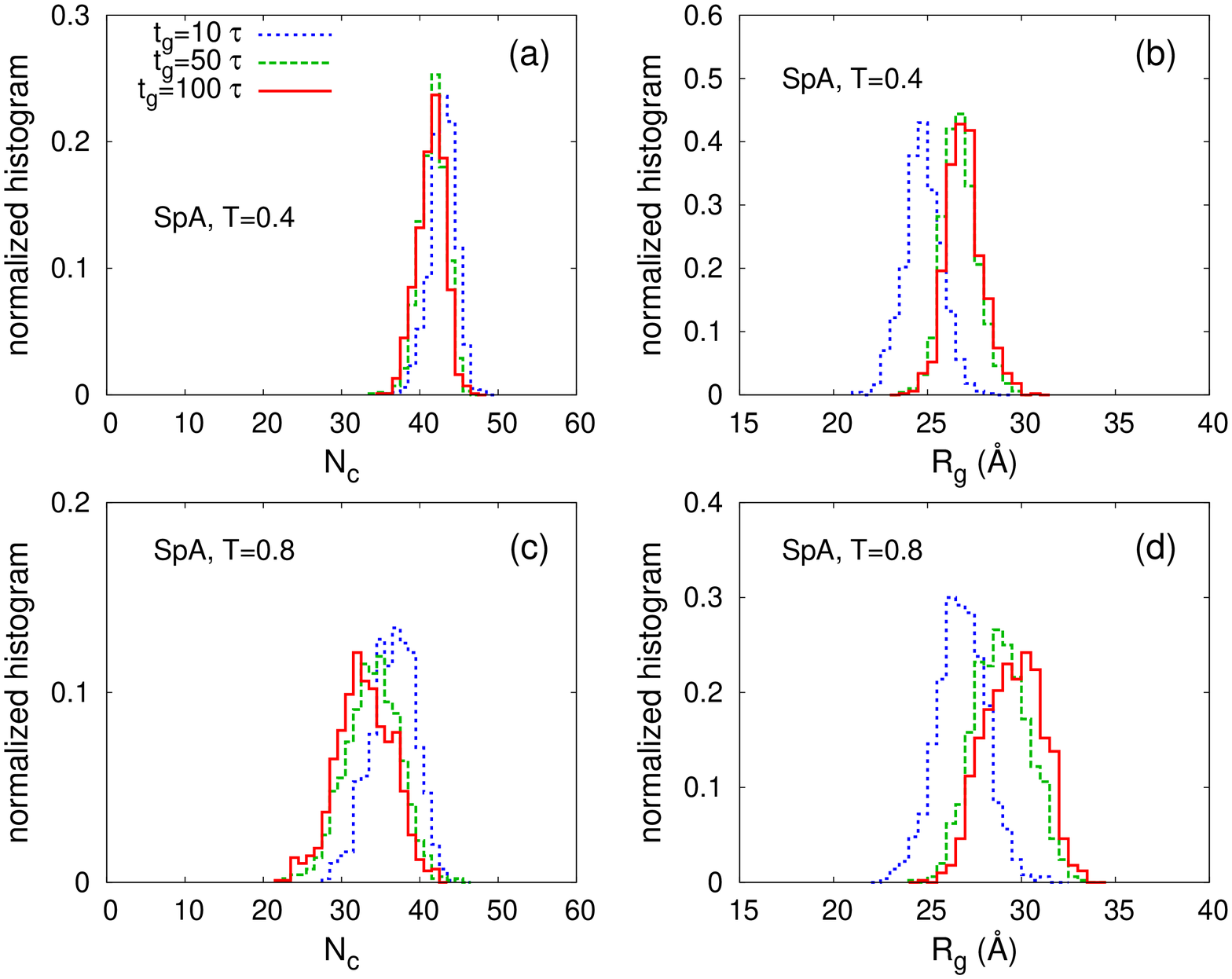}
\caption{
Histograms of conformations of the full length protein at the moment of
complete translation as functions of the number of native
contacts, $N_c$, and the radius of gyration, $R_g$, for protein SpA at
temperatures $T=0.4 \epsilon/k_B$ (a,b) and
$T=0.8 \epsilon/k_B$ (c,d), as indicated. The conformation ensembles are 
generated from 1000 independent growth simulations for each temperature and for
a given growth speed characterized by $t_g$ as the time needed for the chain
elongation of one amino acid. Three different growth speeds are shown for each
temperature as indicated.
}
\end{figure}

\begin{figure}[!ht]
\centering
\includegraphics[width=3in]{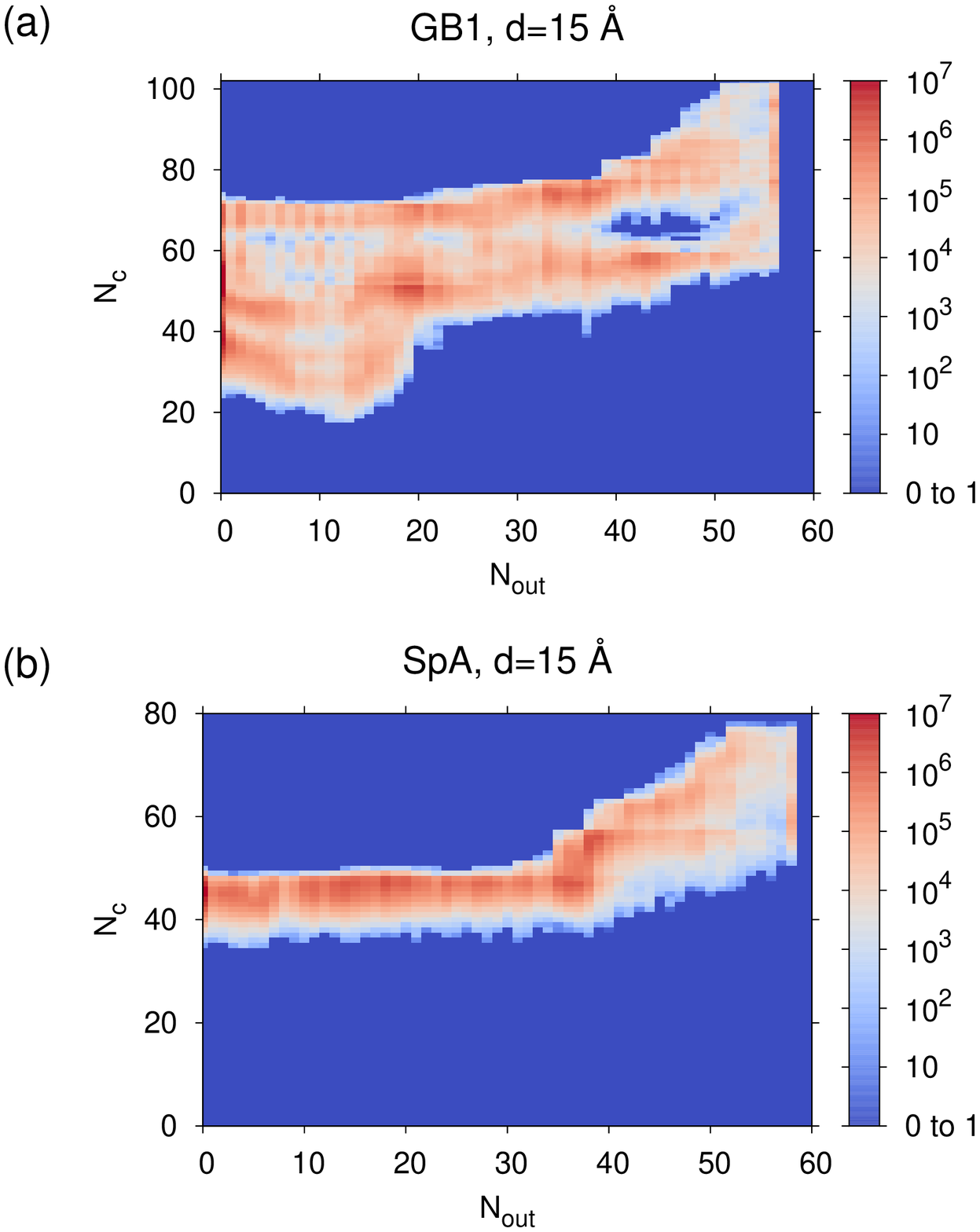}
\caption{
Histogram of conformations obtained during folding and escape of 
protein GB1 (a) and protein SpA (b) at
the tunnel with 4 attractive sites. The histograms were computed from 100
independent folding trajectories at temperature $T=0.4\,\epsilon/k_B$ 
after a translation with $t_g=10\tau$.
}
\end{figure}

\end{document}